# Spin splitting spectroscopy of heavy Quark and antiquarks systems

Hesham Mansour* , Ahmed Gamal and M. Abolmahasen

Physics Department, faculty of science, Cairo University, Giza, Egypt
Corresponding author, Ahmed Gamal.
E-mail: alnabolci2010@gmail.com
* FInstP

## Abstract

Phenomenological potentials describe the quarkonium systems like ($c\bar{c}, b\bar{b}$ and $\bar{b}c$) where they give a good accuracy for the mass spectra. In the present work we extend one of our previous works in the central case by adding spin dependent terms to allow for relativistic corrections. By using such terms, we get better accuracy than previous theoretical calculations. In the present work, the mass spectra of the bound states of heavy quarks $c\bar{c}$, $b\bar{b}$ and $Bc$ mesons are studied within the framework of the non-relativistic Schrödinger's equation. First, we solve Schrödinger's equation by Nikiforov- Uvarov (NU) method. The energy eigenvalues are presented using our new potential. The results obtained are in good agreement with the experimental data and are better than the previous theoretical estimates.

## I. Introduction

In the twenty's century, quarkonium systems have been discovered. Theorists have been trying to explain some aspects of those systems like mass spectra and decay modes properties, etc. [1-5].Some of them used lattice quantum chromo dynamics view[6-12]**,**effective field theory[13], relativistic potential models[14,15], semi-relativistic potential models [16] and non-relativistic potential models[17-19] which have shared in common the Coulomb and linear potentials. There are other groups which use confinement power potential $r^n$[20-22], the Bethe-Salpeter approach[23-25]. In the present work we use mixed potential; nonrelativistic potential models (Coulomb + linear) and confinement power potentials plus spin dependent splitting terms as a correction. Schrödinger's equation is solved by the Nikiforov-Uvarov(NU) method [26-34], which gives asymptotic expressions for the eigenfunctions and eigenvalues of the Schrödinger's equation.

## II. Methodology

In the quarkonium system which deals with quark and antiquark interaction in the center of mass frame, the masses of the quark and antiquark are bigger than chromo dynamics scaling i.e. $M_{q,\bar{q}} \gg \Lambda_{QCD}$. So, this allows for non-relativistic treatment and is considered as heavy bound systems. Using Schrödinger equation of the two-body system in a spherical symmetric potential one obtains:

$$\frac{d^2Q(r)}{dr^2} + \left[\frac{2\mu}{\hbar^2}(E - V_{tot}(r)) - \frac{l(l+1)}{r^2}\right]Q(r) = 0 \tag{1}$$

Where $\mu$ is reduced mass, $E$ is energy eigenvalue, $l$ is orbital quantum number, $V_{tot}(r)$ is total potential of the system, $Q(r) = rR(r)$ and $R(r)$ is a radial wave function solution of Schrödinger's equation.
Our radial potential is taken as:

$$V(r) = \frac{-b}{r} + ar + dr^2 + pr^4 \tag{2}$$

Also, we use spin dependent splitting terms: spin-spin interaction, spin-orbital interaction and tensor interaction respectively[35-43].

$$V_{S-l-T}(r) = V_{S-S}(r) + V_{S-l}(r) + V_T(r) \tag{3}$$

Where

$$V_{S-S}(r) = \frac{2}{3m_q m_{\bar{q}}} \nabla^2 V_V(r) [\vec{S_q} \cdot \vec{S_{\bar{q}}}] \tag{4}$$

$$V_{S-l}(r) = \frac{1}{2m_q m_{\bar{q}} r}\left[3\frac{dV_V(r)}{dr} - \frac{dV_s(r)}{dr}\right][\vec{L} \cdot \vec{S}] \tag{5}$$

$$V_T(r) = \frac{1}{12m_q m_{\bar{q}}}\left[\frac{1}{r}\frac{dV_V(r)}{dr} - \frac{d^2V_V(r)}{dr^2}\right]\left[6\left(\vec{S_q} \cdot \frac{\vec{r}}{|r|}\right)\left(\vec{S_{\bar{q}}} \cdot \frac{\vec{r}}{|r|}\right) - 2\vec{S_q} \cdot \vec{S_{\bar{q}}}\right] \tag{6}$$

$V_V$ is a vector potential term and $V_s$ is a scalar potential term.
So, the total potential becomes

$$V_{tot}(r) = V(r) + V_{S-l-T}(r)$$

$$V_{tot}(r) = \frac{4a_v}{3m_q m_{\bar{q}} r} v(ss) + \frac{v(ls)}{m_q m_{\bar{q}}}\left[\frac{3a_v - a_s}{2r} + \frac{3b}{2r^3} - d - 2pr^2\right] + \frac{v(T)}{12m_q m_{\bar{q}}}\left[\frac{a_v}{r} + \frac{3b}{r^3}\right]$$
$$-\frac{b}{r} + ar + dr^2 + pr^4 \tag{7}$$

Where $v(ss) = [\vec{S_q} \cdot \vec{S_{\bar{q}}}]$ , $v(ls) = [\vec{L} \cdot \vec{S}]$ , $v(T) = -2\left[\vec{S_q} \cdot \vec{S_{\bar{q}}} - 3\left(\vec{S_q} \cdot \frac{\vec{r}}{|r|}\right)\left(\vec{S_{\bar{q}}} \cdot \frac{\vec{r}}{|r|}\right)\right]$

$$a_s + a_v = a$$

By substituting in equation (1), we get

$$\frac{d^2Q}{dr^2} + \left[\varepsilon - \frac{B}{r^3} - \frac{l(l+1)}{r^2} - \frac{C}{r} - Ar - Fr^2 - Pr^4\right]Q = 0 \qquad (8)$$

Where
In natural units

$$\varepsilon = 2\mu E + \frac{2\mu dv(ls)}{m_q m_{\bar{q}}} \quad , \quad B = \frac{\mu b[6v(ls) + v(t)]}{2m_q m_{\bar{q}}} \qquad (9)$$

$$A = 2\mu a \quad , \quad F = \frac{2\mu}{m_q m_{\bar{q}}}[-2pv(ls) + dm_q m_{\bar{q}}] \quad (10)$$

$$C = \frac{\mu[16a_v v(ss) + 6(3a_v - a_s)v(ls) + a_v v(T) - 12m_q m_{\bar{q}} b]}{6m_q m_{\bar{q}}} \quad , \quad P = 2\mu p \qquad (11)$$

let $x = \frac{1}{r}$ and by substituting in equation (8), we obtain

$$\frac{d^2Q}{dx^2} + \frac{2}{x}\frac{dQ}{dx} + \frac{1}{x^2}\left[\frac{\varepsilon}{x^2} - Bx - \frac{C}{x} - l(l+1) - \frac{A}{x^3} - \frac{F}{x^4} - \frac{P}{x^6}\right]Q = 0 \qquad (12)$$

In equation (12), one can use the Nikiforov-Uvarov method (NU) to get eigenvalue and eigenfunction equations.

Due to the singularity point in equation (12), put $y + \delta = x$, and using the Taylor's series to expand to second order terms, one obtains

$$\frac{d^2Q}{dx^2} + \frac{2}{x}\frac{dQ}{dx} + \frac{1}{x^2}[-q + wx - zx^2]Q = 0 \qquad (13)$$

Where

$$-\frac{6\varepsilon}{\delta^2} + \frac{3C}{\delta} + l(l+1) + \frac{10A}{\delta^3} + \frac{15F}{\delta^4} + \frac{28P}{\delta^6} = q \qquad (14)$$

$$-\frac{8\varepsilon}{\delta^3} - B + \frac{3C}{\delta^2} + \frac{15A}{\delta^4} + \frac{24F}{\delta^5} + \frac{48P}{\delta^7} = w \qquad (15)$$

$$-\frac{3\varepsilon}{\delta^4} + \frac{C}{\delta^3} + \frac{6A}{\delta^5} + \frac{10F}{\delta^6} + \frac{21P}{\delta^8} = z \qquad (16)$$

We get

$$z = \left[\frac{w}{2n + 1 + 2\sqrt{q + \frac{1}{4}}}\right]^2 \qquad (17)$$

By substituting equations (14-16) in equation (17) and arrange it, we get

$$\varepsilon = \frac{C\delta}{3} + \frac{2A}{\delta} + \frac{10F}{3\delta^2} + \frac{7P}{\delta^4}$$

$$-\frac{1}{3}\left[\frac{3C + \frac{15A}{\delta^2} + \frac{24F}{\delta^3} + \frac{48P}{\delta^5} - \frac{8\varepsilon}{\delta} - B\delta^2}{2n + 1 + 2\sqrt{\left(l+\frac{1}{2}\right)^2 + \frac{3C}{\delta} - \frac{6\varepsilon}{\delta^2} + \frac{10A}{\delta^3} + \frac{15F}{\delta^4} + \frac{28P}{\delta^6}}}\right]^2 \quad (18)$$

We substitute equations (9-11) into equation (18), to obtain the energy eigenvalue equation

$$E = \aleph - \frac{\mu}{6}\left[\frac{\xi}{2n + 1 + 2\sqrt{\left(l+\frac{1}{2}\right)^2 + \mu\eta}}\right]^2 \quad (19)$$

Where $\delta = \frac{1}{r_0}$, $\aleph = \frac{[16a_v v(ss) + 6(3a_v - a_S)v(ls) + a_v v(t) - 12m_q m_{\bar{q}} b]}{36 m_q m_{\bar{q}} r_0} + 2ar_0 +$
$\frac{10}{3m_q m_{\bar{q}}}[-2pv(ls) + dm_q m_{\bar{q}}]r_0^2 + 7pr_0^4 - \frac{dv(ls)}{m_q m_{\bar{q}}}$

$$\xi = \frac{[16a_v v(ss) + 6(3a_v - a_S)v(ls) + a_v v(t) - 12m_q m_{\bar{q}} b]}{2m_q m_{\bar{q}}} + 30ar_0^2$$
$$+ \frac{48}{m_q m_{\bar{q}}}[-2pv(ls) + dm_q m_{\bar{q}}]r_0^3 + 96pr_0^5 - 8r_0\left[2E + \frac{2dv(ls)}{m_q m_{\bar{q}}}\right]$$
$$- \frac{b[6v(ls) + v(t)]}{2m_q m_{\bar{q}} r_0^2}$$

$$\eta = \left[\frac{[16a_v v(ss) + 6(3a_v - a_S)v(ls) + a_v v(t) - 12m_q m_{\bar{q}} b]r_0}{2m_q m_{\bar{q}}} + 20ar_0^3\right.$$
$$\left. + \frac{30}{m_q m_{\bar{q}}}[-2pv(ls) + dm_q m_{\bar{q}}]r_0^4 + 56pr_0^6 - 6r_0^2\left[2E + \frac{2dv(ls)}{m_q m_{\bar{q}}}\right]\right]$$

Knowing that

$$M(q\bar{q}) = E + m_q + m_{\bar{q}} \quad \rightarrow \quad E = M(q\bar{q}) - (m_q + m_{\bar{q}}) \quad (20)$$

So, the mass spectra equation becomes,

$$M(q\bar{q}) = \aleph - \frac{\mu}{6}\left[\frac{\xi}{2n + 1 + 2\sqrt{\left(l+\frac{1}{2}\right)^2 + \mu\eta}}\right]^2 + m_q + m_{\bar{q}} \quad (21)$$

The eigenfunction equation is

$$Q(r) = N_{nls} r^{\frac{1}{2}-\sqrt{\left(q+\frac{1}{4}\right)}} e^{-\frac{\sqrt{z}}{r}} L_n^{2\sqrt{\left(q+\frac{1}{4}\right)}} \left(\frac{2\sqrt{z}}{r}\right) \quad (22)$$

Where $L_n^{2\sqrt{\left(q+\frac{1}{4}\right)}}\left(\frac{2\sqrt{z}}{r}\right)$ is the Rodrigues's formula of the associated Laguerre polynomial and $N_{nls}$ is a normalization constant.

So, the radial wave function solution of Schrödinger's equation is given by.

$$R(r) = N_{nls} r^{-\frac{1}{2}-\sqrt{\left(q+\frac{1}{4}\right)}} e^{-\frac{\sqrt{z}}{r}} L_n^{2\sqrt{\left(q+\frac{1}{4}\right)}} \left(\frac{2\sqrt{z}}{r}\right) \quad (23)$$

The energy eigenvalue equation (19) has spin-orbital-tensor coefficients $v(ss), v(sl), v(T)$ and those can be given from references[35-37]. Also, it has potential parameters $(a_s, a_v, b, d, p)$ and $r_0$ due to the expansion, so we have six parameters of the eigenvalue equation which can be obtained from the experimental data[44] by best fitting as shown in appendix.

## III. Numerical Results and Discussions

In table (1), potential parameters are shown for each system. It is noticed that the values of these parameters are different for different systems and this is due to the properties of those systems like energy scale, decay mode...etc. We use spectroscopic notation for the levels ($n^{2S+1}L_J$).
S is total spin of the system, L is the orbital quantum number, n is the principal quantum number, J is the total (orbital + spin) quantum number.
By using equation (21) and table (1), we get the mass spectra of different quantum states as shown in the tables (2-7). Previously, we used the phenomenological potential in equation (2) without spin dependent corrections (central dependent potential)[45]. The results obtained were good in comparison with the experimental data.

Table 1. Parameter values of each system

| Variables / Systems | $m_q$ | $m_{\bar{q}}$ | $r_0$ | as | av | b | d | p |
|---|---|---|---|---|---|---|---|---|
| units | GeV | GeV | GeV$^{-1}$ | GeV$^2$ | GeV$^2$ | -- | GeV$^3$ | GeV$^5$ |
| $c\bar{c}$ system | 1.317 | 1.317 | 12.82 | -0.0796 | -0.7349 | 0.009 | 0.10686 | -0.000184 |
| $b\bar{b}$ system | 4.584 | 4.584 | 7.23795 | 0.0505 | 2.5771 | 15.02 | −0.07969 | −0.00102 |
| $\bar{b}c$ system | 4.584 | 1.317 | 11.434 | −2.5549 | 1.5369 | 0.039 | 0.11453 | −0.00018438 |

Table 2. Charmonia mass spectrum of S and P-wave in GeV

| Level | Present work | [24] | [46] | [37] | [47] | [48] | [15] | [49] | [40] | [50] | [51] | Exp. [44] |
|---|---|---|---|---|---|---|---|---|---|---|---|---|
| $1^1S_0$ | 3.033 | 2.93 | 2.981 | 2.984 | 2.989 | 2.979 | 2.980 | 2.980 | 2.982 | 3.088 | 2.979 | 2.984(49) |
| $1^3S_1$ | 3.126 | 3.11 | 3.096 | 3.097 | 3.094 | 3.097 | 3.097 | 3.097 | 3.090 | 3.168 | 3.096 | 3.097(29) |
| $2^1S_0$ | 3.666 | 3.68 | 3.635 | 3.637 | 3.602 | 3.623 | 3.597 | 3.633 | 3.630 | 3.669 | 3.600 | 3.639(27) |
| $2^3S_1$ | 3.701 | 3.68 | 3.685 | 3.679 | 3.681 | 3.673 | 3.685 | 3.690 | 3.672 | 3.707 | 3.680 | 3.686(15) |
| $3^1S_0$ | 4.158 | -- | 3.989 | 4.004 | 4.058 | 3.991 | 4.014 | 3.992 | 4.043 | 4.067 | 4.011 | -- |
| $3^3S_1$ | 4.055 | 3.80 | 4.039 | 4.030 | 4.129 | 4.022 | 4.095 | 4.030 | 4.072 | 4.094 | 4.077 | 4.039(16) |
| $4^1S_0$ | 4.415 | -- | 4.401 | 4.264 | 4.448 | 4.250 | 4.433 | 4.244 | 4.384 | 4.398 | 4.397 | -- |
| $4^3S_1$ | 4.415 | -- | 4.427 | 4.281 | 4.514 | 4.273 | 4.477 | 4.273 | 4.406 | 4.420 | 4.454 | 4.421(6) |
| $5^1S_0$ | 4.607 | -- | 4.811 | 4.459 | 4.799 | 4.446 | -- | 4.440 | -- | -- | -- | -- |
| $5^3S_1$ | 4.585 | -- | 4.837 | 4.472 | 4.863 | 4.463 | -- | 4.464 | -- | -- | -- | -- |
| $6^1S_0$ | 4.754 | -- | 5.155 | -- | 5.124 | 4.595 | -- | 4.601 | -- | -- | -- | -- |
| $6^3S_1$ | 4.733 | -- | 5.167 | -- | 5.185 | 4.608 | -- | 4.621 | -- | -- | -- | -- |
| $1^3P_0$ | 3.407 | 3.32 | 3.413 | 3.415 | 3.428 | 3.433 | 3.416 | 3.392 | 3.424 | 3.448 | 3.488 | 3.415(8) |
| $1^3P_1$ | 3.487 | 3.49 | 3.511 | 3.521 | 3.468 | 3.510 | 3.508 | 3.491 | 3.505 | 3.520 | 3.514 | 3.511(24) |
| $1^1P_1$ | 3.502 | 3.43 | 3.525 | 3.526 | 3.470 | 3.519 | 3.527 | 3.524 | 3.516 | 3.536 | 3.536 | 3.525(23) |
| $1^3P_2$ | 3.522 | 3.55 | 3.555 | 3.553 | 3.480 | 3.556 | 3.558 | 3.570 | 3.556 | 3.564 | 3.565 | 3.556(33) |
| $2^3P_0$ | 3.899 | 3.83 | 3.870 | 3.848 | 3.897 | 3.842 | 3.844 | 3.845 | 3.852 | 3.870 | 3.947 | 3.918(19) |
| $2^3P_1$ | 3.786 | 3.67 | 3.906 | 3.914 | 3.938 | 3.901 | 3.940 | 3.902 | 3.925 | 3.934 | 3.972 | -- |
| $2^1P_1$ | 3.821 | 3.75 | 3.926 | 3.916 | 3.943 | 3.908 | 3.960 | 3.922 | 3.934 | 3.950 | 3.996 | -- |
| $2^3P_2$ | 3.905 | -- | 3.949 | 3.937 | 3.955 | 3.937 | 3.994 | 3.949 | 3.972 | 3.976 | 4.021 | 3.927(21) |
| $3^3P_0$ | 4.120 | -- | 4.301 | 4.146 | 4.296 | 4.131 | -- | 4.192 | 4.202 | 4.214 | -- | -- |
| $3^3P_1$ | 4.123 | 3.91 | 4.319 | 4.192 | 4.338 | 4.178 | -- | 4.178 | 4.271 | 4.275 | -- | -- |

| Level | | | | | | | | | | | |
|---|---|---|---|---|---|---|---|---|---|---|---|
| $3^1P_1$ | 4.164 | -- | 4.337 | 4.193 | 4.344 | 4.184 | -- | 4.137 | 4.279 | 4.291 | -- | -- |
| $3^3P_2$ | 4.144 | -- | 4.354 | 4.211 | 4.358 | 4.208 | -- | 4.212 | 4.317 | 4.316 | -- | -- |
| $4^3P_0$ | 4.362 | -- | 4.698 | -- | 4.653 | -- | -- | -- | -- | -- | -- | -- |
| $4^3P_1$ | 4.373 | -- | 4.728 | -- | 4.696 | -- | -- | -- | -- | -- | -- | -- |
| $4^1P_1$ | 4.42 | -- | 4.744 | -- | 4.704 | -- | -- | -- | -- | -- | -- | -- |
| $4^3P_2$ | 4.411 | -- | 4.763 | -- | 4.718 | -- | -- | -- | -- | -- | -- | -- |
| $5^3P_0$ | 4.543 | -- | -- | -- | 4.983 | -- | -- | -- | -- | -- | -- | -- |
| $5^3P_1$ | 4.561 | -- | -- | -- | 5.026 | -- | -- | -- | -- | -- | -- | -- |
| $5^1P_1$ | 4.611 | -- | -- | -- | 5.034 | -- | -- | -- | -- | -- | -- | -- |

Table 3. Charmonia mass spectrum of D and F-wave in GeV

| Level | Present work | [47] | [46] | [37] | [24] | [48] | [15] | [49] | [40] | [50] | [51] |
|---|---|---|---|---|---|---|---|---|---|---|---|
| $1^3D_3$ | 3.307 | 3.755 | 3.813 | 3.808 | 3.869 | 3.799 | 3.831 | 3.844 | 3.806 | 3.809 | 3.798 |
| $1^1D_2$ | 3.376 | 3.765 | 3.807 | 3.805 | 3.739 | 3.796 | 3.824 | 3.802 | 3.799 | 3.803 | 3.796 |
| $1^3D_2$ | 3.348 | 3.772 | 3.795 | 3.807 | 3.550 | 3.798 | 3.824 | 3.788 | 3.800 | 3.804 | 3.794 |
| $1^3D_1$ | 3.374 | 3.775 | 3.783 | 3.792 | -- | 3.787 | 3.804 | 3.729 | 3.785 | 3.789 | 3.792 |
| $2^3D_3$ | 3.797 | 4.176 | 4.220 | 4.112 | 3.806 | 4.103 | 4.202 | 4.132 | 4.167 | 4.167 | 4.425 |
| $2^1D_2$ | 3.836 | 4.182 | 4.196 | 4.108 | -- | 4.099 | 4.191 | 4.105 | 4.158 | 4.158 | 4.224 |
| $2^3D_2$ | 3.801 | 4.188 | 4.190 | 4.109 | -- | 4.100 | 4.189 | 4.095 | 4.158 | 4.159 | 4.223 |
| $2^3D_1$ | 3.800 | 4.188 | 4.105 | 4.095 | -- | 4.089 | 4.164 | 4.057 | 4.142 | 4.143 | 4.222 |
| $3^3D_3$ | 4.163 | 4.549 | 4.574 | 4.340 | -- | 4.331 | -- | 4.351 | -- | -- | -- |
| $3^1D_2$ | 4.176 | 4.553 | 3.549 | 4.336 | -- | 4.326 | -- | 4.330 | -- | -- | -- |
| $3^3D_2$ | 4.135 | 4.557 | 4.544 | 4.337 | -- | 4.327 | -- | 4.322 | -- | -- | -- |

| Level | Present | [52] | [46] | [37] | [24] | [53] | [15] | [49] | [54] | Exp. [44] |
|---|---|---|---|---|---|---|---|---|---|---|
| $3^3D_1$ | 4.113 | 4.555 | 4.507 | 4.324 | -- | 4.317 | -- | 4.293 | -- | -- | -- |
| $4^3D_3$ | 4.436 | 4.890 | 4.920 | -- | -- | -- | -- | 4.526 | -- | -- | -- |
| $4^1D_2$ | 4.429 | 4.892 | 4.898 | -- | -- | -- | -- | 4.509 | -- | -- | -- |
| $4^3D_2$ | 4.383 | 4.896 | 4.896 | -- | -- | -- | -- | 4.504 | -- | -- | -- |
| $4^3D_1$ | 4.345 | 4.891 | 4.857 | -- | -- | -- | -- | 4.480 | -- | -- | -- |
| $1^3F_2$ | 3.403 | 3.990 | 4.041 | -- | -- | -- | 4.068 | -- | 4.029 | -- | -- |
| $1^3F_3$ | 3.375 | 4.012 | 4.068 | -- | 3.999 | -- | 4.070 | -- | 4.029 | -- | -- |
| $1^1F_3$ | 3.403 | 4.017 | 4.071 | -- | 4.037 | -- | 4.066 | -- | 4.026 | -- | -- |
| $1^3F_4$ | 3.315 | 4.036 | 4.093 | -- | -- | -- | 4.062 | -- | 4.021 | -- | -- |
| $2^3F_2$ | 3.812 | 4.378 | 4.361 | -- | -- | -- | -- | -- | 4.351 | -- | -- |
| $2^3F_3$ | 3.823 | 4.396 | 4.400 | -- | -- | -- | -- | -- | 3.352 | -- | -- |
| $2^1F_3$ | 3.858 | 4.400 | 4.406 | -- | -- | -- | -- | -- | 4.350 | -- | -- |
| $2^3F_4$ | 3.814 | 4.415 | 4.434 | -- | -- | -- | -- | -- | 4.348 | -- | -- |
| $3^3F_2$ | 4.111 | 4.730 | -- | -- | -- | -- | -- | -- | -- | -- | -- |
| $3^3F_3$ | 4.152 | 4.746 | -- | -- | -- | -- | -- | -- | -- | -- | -- |
| $3^1F_3$ | 4.194 | 4.749 | -- | -- | -- | -- | -- | -- | -- | -- | -- |
| $3^3F_4$ | 4.186 | 4.761 | -- | -- | -- | -- | -- | -- | -- | -- | -- |

Table 4. Bottomonia mass spectrum of S and P-wave in GeV

| | work | | | | | | | | | |
|---|---|---|---|---|---|---|---|---|---|---|
| $1^1S_0$ | 9.472 | 9.402 | 9.398 | 9.390 | 9.414 | 9.389 | 9.393 | 9.392 | 9.455 | 9.398(74) |
| $1^3S_1$ | 9.525 | 9.465 | 9.460 | 9.460 | 9.490 | 9.460 | 9.460 | 9.460 | 9.502 | 9.460(65) |
| $2^1S_0$ | 10.028 | 9.976 | 9.990 | 9.990 | 9.987 | 9.987 | 9.987 | 9.991 | 9.990 | 9.999(28) |
| $2^3S_1$ | 10.049 | 10.003 | 10.023 | 10.015 | 10.089 | 10.016 | 10.023 | 10.024 | 10.015 | 10.023(26) |
| $3^1S_0$ | 10.36 | 10.336 | 10.329 | 10.326 | -- | 10.330 | 10.345 | 10.323 | 10.330 | -- |
| $3^3S_1$ | 10.371 | 10.354 | 10.355 | 10.343 | 10.327 | 10.351 | 10.364 | 10.346 | 10.349 | 10.355(16) |
| $4^1S_0$ | 10.592 | 10.523 | 10.573 | 10.584 | -- | 10.595 | 10.623 | 10.558 | -- | -- |
| $4^3S_1$ | 10.598 | 10.635 | 10.586 | 10.597 | -- | 10.611 | 10.643 | 10.575 | 10.607 | 10.579(19) |
| $5^1S_0$ | 10.79 | 10.869 | 10.851 | 10.800 | -- | 10.817 | -- | 10.741 | -- | -- |
| $5^3S_1$ | 10.87 | 10.878 | 10.869 | 10.811 | -- | 10.831 | -- | 10.755 | 10.818 | 10.876(6) |
| $6^1S_0$ | 10.961 | 11.097 | 11.061 | 10.997 | -- | 11.011 | -- | 10.892 | -- | -- |
| $6^3S_1$ | 11.022 | 11.102 | 11.088 | 10.988 | -- | 10.988 | -- | 10.904 | 10.995 | 11.019(3) |
| $1^3P_0$ | 9.84 | 9.847 | 9.859 | 9.864 | 9.815 | 9.865 | 9.861 | 9.862 | 9.855 | 9.859(19) |
| $1^3P_1$ | 9.875 | 9.876 | 9.892 | 9.903 | 9.842 | 9.897 | 9.891 | 9.888 | 9.874 | 9.893(18) |
| $1^1P_1$ | 9.884 | 9.882 | 9.900 | 9.909 | 9.806 | 9.903 | 9.900 | 9.896 | 9.879 | 9.899(15) |
| $1^3P_2$ | 9.903 | 9.897 | 9.912 | 9.921 | 9.906 | 9.918 | 9.912 | 9.908 | 9.886 | 9.912(9) |
| $2^3P_0$ | 10.202 | 10.226 | 10.233 | 10.220 | 10.254 | 10.226 | 10.230 | 10.241 | 10.221 | 10.232(30) |
| $2^3P_1$ | 10.229 | 10.246 | 10.255 | 10.249 | 10.120 | 10.251 | 10.255 | 10.256 | 10.236 | 10.255(26) |

| Level | Present work | [52] | [46] | [37] | [24] | [53] | [15] | [49] | [54] | Exp. |
|---|---|---|---|---|---|---|---|---|---|---|
| $2^1P_1$ | 10.237 | 10.250 | 10.260 | 10.254 | 10.154 | 10.256 | 10.262 | 10.261 | 10.240 | 10.260(23) |
| $2^3P_2$ | 10.254 | 10.261 | 10.268 | 10.264 | -- | 10.269 | 10.271 | 10.268 | 10.246 | 10.269(15) |
| $3^3P_0$ | 10.299 | 10.552 | 10.521 | 10.490 | -- | 10.502 | -- | 10.511 | 10.500 | -- |
| $3^3P_1$ | 10.339 | 10.538 | 10.541 | 10.515 | 10.303 | 10.524 | -- | 10.507 | 10.513 | -- |
| $3^1P_1$ | 10.362 | 10.541 | 10.544 | 10.519 | -- | 10.529 | -- | 10.497 | 10.516 | -- |
| $3^3P_2$ | 10.406 | 10.550 | 10.550 | 10.528 | -- | 10.540 | -- | 10.516 | 10.521 | -- |
| $4^3P_0$ | 10.532 | 10.775 | 10.781 | -- | -- | 10.732 | -- | -- | -- | -- |
| $4^3P_1$ | 10.571 | 10.788 | 10.802 | -- | -- | 10.753 | -- | -- | -- | -- |
| $4^1P_1$ | 10.594 | 10.790 | 10.804 | -- | -- | 10.757 | -- | -- | -- | -- |
| $4^3P_2$ | 10.637 | 10.798 | 10.812 | -- | -- | 10.767 | -- | -- | -- | -- |
| $5^3P_0$ | 10.731 | 11.004 | -- | -- | -- | 10.933 | -- | -- | -- | -- |
| $5^3P_1$ | 10.769 | 11.014 | -- | -- | -- | 10.951 | -- | -- | -- | -- |
| $5^1P_1$ | 10.792 | 11.016 | -- | -- | -- | 10.955 | -- | -- | -- | -- |

Table 5. Bottomonia mass spectrum of D and F-wave in GeV

| Level | Present work | [52] | [46] | [37] | [24] | [53] | [15] | [49] | [54] | Exp. [44] |
|---|---|---|---|---|---|---|---|---|---|---|
| $1^3D_3$ | 9.849 | 10.115 | 10.166 | 10.157 | 10.232 | 10.156 | 10.163 | 10.177 | 10.127 | -- |

| State | | | | | | | | | | |
|---|---|---|---|---|---|---|---|---|---|---|
| $1^1D_2$ | 9.767 | 10.148 | 10.163 | 10.153 | 10.194 | 10.152 | 10.158 | 10.166 | 10.123 | -- |
| $1^3D_2$ | 10.096 | 10.147 | 10.161 | 10.153 | 10.145 | 10.151 | 10.157 | 10.162 | 10.122 | 10.163(66) |
| $1^3D_1$ | 9.666 | 10.138 | 10.154 | 10.146 | -- | 10.145 | 10.149 | 10.147 | 10.117 | -- |
| $2^3D_3$ | 10.175 | 10.455 | 10.449 | 10.436 | -- | 10.442 | 10.456 | 10.447 | 10.422 | -- |
| $2^1D_2$ | 10.093 | 10.450 | 10.445 | 10.432 | -- | 10.439 | 10.452 | 10.440 | 10.419 | -- |
| $2^3D_2$ | 10.071 | 10.449 | 10.443 | 10.432 | -- | 10.438 | 10.450 | 10.437 | 10.418 | -- |
| $2^3D_1$ | 9.996 | 10.441 | 10.435 | 10.425 | -- | 10.432 | 10.443 | 10.428 | 10.414 | -- |
| $3^3D_3$ | 10.446 | 10.711 | 10.717 | -- | -- | 10.680 | -- | 10.652 | -- | -- |
| $3^1D_2$ | 10.368 | 10.706 | 10.713 | -- | -- | 10.677 | -- | 10.646 | --- | -- |
| $3^3D_2$ | 10.345 | 10.705 | 10.711 | -- | -- | 10.676 | -- | 10.645 | -- | -- |
| $3^3D_1$ | 10.272 | 10.698 | 10.704 | -- | -- | 10.670 | -- | 10.637 | -- | -- |
| $4^3D_3$ | 10.676 | 10.939 | 10.963 | -- | -- | 10.886 | -- | 10.817 | -- | -- |
| $4^1D_2$ | 10.599 | 10.935 | 10.959 | -- | -- | 10.883 | -- | 10.813 | -- | -- |
| $4^3D_2$ | 10.576 | 10.934 | 10.957 | -- | -- | 10.882 | -- | 10.811 | -- | -- |
| $4^3D_1$ | 10.504 | 10.928 | 10.949 | -- | -- | 10.877 | -- | 10.811 | -- | -- |
| $1^3F_2$ | 9.642 | 10.350 | 10.343 | 10.338 | -- | -- | 10.353 | -- | 10.315 | -- |
| $1^3F_3$ | 9.754 | 10.355 | 10.346 | 10.340 | 10.302 | -- | 10.356 | -- | 10.321 | -- |
| $1^1F_3$ | 9.778 | 10.355 | 10.347 | 10.339 | 10.319 | -- | 10.356 | -- | 10.322 | -- |
| $1^3F_4$ | 9.896 | 10.358 | 10.349 | 10.340 | -- | -- | 10.357 | -- | -- | -- |
| $2^3F_2$ | 9.971 | 10.615 | 10.610 | -- | -- | -- | 10.610 | -- | -- | -- |
| $2^3F_3$ | 10.081 | 10.619 | 10.614 | -- | -- | -- | 10.613 | -- | -- | -- |
| $2^1F_3$ | 10.104 | 10.619 | 10.647 | -- | -- | -- | 10.613 | -- | -- | -- |
| $2^3F_4$ | 10.219 | 10.622 | 10.617 | -- | -- | -- | 10.615 | -- | -- | -- |
| $3^3F_2$ | 10.246 | 10.850 | -- | -- | -- | -- | -- | -- | -- | -- |
| $3^3F_3$ | 10.353 | 10.853 | -- | -- | -- | -- | -- | -- | -- | -- |
| $3^1F_3$ | 10.376 | 10.853 | -- | -- | -- | -- | -- | -- | -- | -- |

| Level | Present work | | | | | | |
|---|---|---|---|---|---|---|---|
| $3^3F_4$ | 10.489 | 10.856 | -- | -- | -- | -- | -- | -- | -- | -- |

Table 6. $B_c$ Meson mass spectrum of S and P-wave in GeV

| Level | Present work | [47] | [55] | [46] | [56] | [57] | Exp.[44] |
|---|---|---|---|---|---|---|---|
| $1^1S_0$ | 6.276 | 6.272 | 6.278 | 6.272 | 6.271 | 6.275 | 6.275(0) |
| $1^3S_1$ | 6.313 | 6.321 | 6.331 | 6.333 | 6.338 | 6.314 | -- |
| $2^1S_0$ | 6.841 | 6.864 | 6.863 | 6.842 | 6.855 | 6.838 | 6.842(1) |
| $2^3S_1$ | 6.867 | 6.900 | 6.873 | 6.882 | 6.887 | 6.850 | -- |
| $3^1S_0$ | 7.281 | 7.306 | 7.244 | 7.226 | 7.250 | -- | -- |
| $3^3S_1$ | 7.308 | 7.338 | 7.249 | 7.258 | 7.272 | -- | -- |
| $4^1S_0$ | 7.634 | 7.684 | 7.564 | 7.585 | -- | -- | -- |
| $4^3S_1$ | 7.66 | 7.714 | 7.568 | 7.609 | -- | -- | -- |
| $5^1S_0$ | 7.917 | 8.025 | 7.852 | 7.928 | -- | -- | -- |
| $5^3S_1$ | 7.941 | 8.054 | 7.855 | 7.947 | -- | -- | -- |
| $6^1S_0$ | 8.144 | 8.340 | 8.120 | -- | -- | -- | -- |
| $6^3S_1$ | 8.168 | 8.368 | 8.122 | -- | -- | -- | -- |
| $1^3P_0$ | 6.223 | 6.686 | 6.748 | 6.699 | 6.706 | 6.672 | -- |
| $1^3P_1$ | 6.281 | 6.705 | 6.767 | 6.750 | 6.741 | 6.766 | -- |
| $1^1P_1$ | 6.290 | 6.706 | 6.769 | 6.743 | 6.750 | 6.828 | -- |
| $1^3P_2$ | 6.366 | 6.712 | 6.775 | 6.761 | 6.768 | 6.776 | -- |
| $2^3P_0$ | 6.782 | 7.146 | 7.139 | 7.094 | 7.122 | 6.914 | -- |
| $2^3P_1$ | 6.836 | 7.165 | 7.155 | 7.134 | 7.145 | 7.259 | -- |
| $2^1P_1$ | 6.846 | 7.168 | 7.156 | 7.094 | 7.150 | 7.322 | -- |
| $2^3P_2$ | 6.917 | 7.173 | 7.162 | 7.157 | 7.164 | 7.232 | -- |
| $3^3P_0$ | 7.227 | 7.536 | 7.463 | 7.474 | -- | -- | -- |
| $3^3P_1$ | 7.278 | 7.555 | 7.479 | 7.510 | -- | -- | -- |

| Level | | | | | | |
|---|---|---|---|---|---|---|
| $3^1P_1$ | 7.287 | 7.559 | 7.479 | 7.500 | -- | -- | -- |
| $3^3P_2$ | 7.355 | 7.565 | 7.485 | 7.524 | -- | -- | -- |
| $4^3P_0$ | 7.583 | 7.885 | -- | 7.817 | -- | -- | -- |
| $4^3P_1$ | 7.631 | 7.905 | -- | 7.853 | -- | -- | -- |
| $4^1P_1$ | 7.640 | 7.908 | -- | 7.844 | -- | -- | -- |
| $4^3P_2$ | 7.704 | 7.915 | -- | 7.867 | -- | -- | -- |
| $5^3P_0$ | 7.867 | 8.207 | -- | -- | -- | -- | -- |
| $5^3P_1$ | 7.913 | 8.226 | -- | -- | -- | -- | -- |
| $5^1P_1$ | 7.922 | 8.230 | -- | -- | -- | -- | -- |

Table 7. $B_c$ Meson mass spectrum of D and F-wave in GeV

| Level | Present work | [47] | [55] | [46] | [56] | [57] |
|---|---|---|---|---|---|---|
| $1^3D_3$ | 6.429 | 6.990 | 7.026 | 7.029 | 7.045 | 6.980 |
| $1^1D_2$ | 6.308 | 6.994 | 7.035 | 7.026 | 7.041 | 7.009 |
| $1^3D_2$ | 6.299 | 6.997 | 7.025 | 7.025 | 7.036 | 7.154 |
| $1^3D_1$ | 6.200 | 6.998 | 7.030 | 7.021 | 7.028 | 7.078 |
| $2^3D_3$ | 6.975 | 7.399 | 7.363 | 7.405 | -- | -- |
| $2^1D_2$ | 6.861 | 7.401 | 7.370 | 7.400 | -- | -- |
| $2^3D_2$ | 6.852 | 7.403 | 7.361 | 7,399 | -- | -- |
| $2^3D_1$ | 6.759 | 7.403 | 7.365 | 7.392 | -- | -- |
| $3^3D_3$ | 7.409 | 7.761 | -- | 7.750 | -- | -- |
| $3^1D_2$ | 7.3017 | 7.762 | -- | 7.743 | -- | -- |
| $3^3D_2$ | 7.292 | 7.764 | -- | 7.741 | -- | -- |
| $3^3D_1$ | 7.205 | 7.762 | -- | 7.732 | -- | -- |
| $4^3D_3$ | 7.754 | 8.092 | -- | -- | -- | -- |
| $4^1D_2$ | 7.653 | 8.093 | -- | -- | -- | -- |
| $4^3D_2$ | 7.643 | 8.094 | -- | -- | -- | -- |
| $4^3D_1$ | 7.56 | 8.091 | -- | -- | -- | -- |
| $1^3F_2$ | 6.183 | 7.234 | -- | 7.273 | 7.269 | -- |
| $1^3F_3$ | 6.326 | 7.242 | -- | 7.269 | 7.276 | -- |
| $1^1F_3$ | 6.335 | 7.241 | -- | 7.268 | 7.266 | -- |
| $1^3F_4$ | 6.501 | 7.244 | -- | 7.277 | 7.271 | -- |
| $2^3F_2$ | 6.741 | 7.607 | -- | 7.618 | -- | -- |
| $2^3F_3$ | 6.876 | 7.615 | -- | 7.616 | -- | -- |
| $2^1F_3$ | 6.885 | 7.614 | -- | 7.615 | -- | -- |
| $2^3F_4$ | 7.041 | 7.617 | -- | 7.617 | -- | -- |
| $3^3F_2$ | 7.186 | 7.946 | -- | -- | -- | -- |
| $3^3F_3$ | 7.314 | 7.954 | -- | -- | -- | -- |
| $3^1F_3$ | 7.323 | 7.953 | -- | -- | -- | -- |
| $3^3F_4$ | 7.470 | 7.956 | -- | -- | -- | -- |

## IV. CONCLUSIONS

The above tables show that, spin dependent terms are important factors to give a better accuracy and complete quantitative description of the quarkonium systems for the cases where experimental values are available. The theoretical work agrees with the experimental data. This shows also that the Nikiforov-Uvarov method is a good method to get the energy eigenvalues for the meson spectra. The results are even better than other previous works.

## References

[1] E. Eichten, S. Godfrey, H. Mahlke, and J. L. Rosner, "Quarkonia and their transitions," *Rev. Mod. Phys.*, vol. 80, no. 3, pp. 1–80, 2008, doi: 10.1103/RevModPhys.80.1161.

[2] N. Brambilla *et al.*, "Heavy Quarkonium Physics," no. June, 2004.

[3] N. Brambilla *et al.*, "Heavy quarkonium: Progress, puzzles, and opportunities," *Eur. Phys. J. C*, vol. 71, no. 2, pp. 1–178, 2011, doi: 10.1140/epjc/s10052-010-1534-9.

[4] A. Andronic *et al.*, "Heavy-flavour and quarkonium production in the LHC era: from proton–proton to heavy-ion collisions," *Eur. Phys. J. C*, vol. 76, no. 3, 2016, doi: 10.1140/epjc/s10052-015-3819-5.

[5] S. Godfrey and S. L. Olsen, "The Exotic XYZ Charmonium-Like Mesons," *Annu. Rev. Nucl. Part. Sci.*, vol. 58, no. 1, pp. 51–73, 2008, doi: 10.1146/annurev.nucl.58.110707.171145.

[6] J. J. Dudek, R. G. Edwards, N. Mathur, and D. G. Richards, "Charmonium excited state spectrum in lattice QCD," *Phys. Rev. D - Part. Fields, Gravit. Cosmol.*, vol. 77, no. 3, 2008, doi: 10.1103/PhysRevD.77.034501.

[7] T. Burch *et al.*, "Quarkonium mass splittings in three-flavor lattice QCD," *Phys. Rev. D - Part. Fields, Gravit. Cosmol.*, vol. 81, no. 3, pp. 1–21, 2010, doi: 10.1103/PhysRevD.81.034508.

[8] T. Liu, A. A. Penin, and A. Rayyan, "Coulomb artifacts and bottomonium hyperfine splitting in lattice NRQCD," *J. High Energy Phys.*, vol. 2017, no. 2, 2017, doi: 10.1007/JHEP02(2017)084.

[9] S. Meinel, "Bottomonium spectrum from lattice QCD with 2+1 flavors of domain wall fermions," *Phys. Rev. D - Part. Fields, Gravit. Cosmol.*, vol. 79, no. 9, pp. 1–11, 2009, doi: 10.1103/PhysRevD.79.094501.

[10] K. Nochi, T. Kawanai, and S. Sasaki, "Bethe-Salpeter wave functions of $\eta_c$(2S) and $\psi$(2S) states from full lattice QCD," *Phys. Rev. D*, vol. 94, no. 11, pp. 1–12, 2016, doi: 10.1103/PhysRevD.94.114514.

[11] C. McNeile, C. T. H. Davies, E. Follana, K. Hornbostel, and G. P. Lepage, "Heavy meson masses and decay constants from relativistic heavy quarks in full lattice QCD," *Phys. Rev. D - Part. Fields, Gravit. Cosmol.*, vol. 86, no. 7, p. 074503, Oct. 2012, doi: 10.1103/PhysRevD.86.074503.

[12] J. O. Daldrop, C. T. H. Davies, and R. J. Dowdall, "Prediction of the bottomonium D-wave spectrum from full lattice QCD," *Phys. Rev. Lett.*, vol. 108, no. 10, pp. 4–7, 2012, doi: 10.1103/PhysRevLett.108.102003.

[13] M. Neubert, "Heavy-quark symmetry," *Phys. Rep.*, vol. 245, no. 5–6, pp. 259–395, 1994, doi: 10.1016/0370-1573(94)90091-4.

[14] K. M. Maung, D. E. Kahana, and J. W. Norbury, "Solution of two-body relativistic bound-state equations with confining plus coulomb interactions," *Phys. Rev. D*, vol. 47, no. 3, pp. 1182–1189, 1993, doi: 10.1103/PhysRevD.47.1182.

[15] S. F. Radford and W. W. Repko, "Hyperfine splittings in the bb- system," *Nucl. Phys. A*, vol. 865, no. 1, pp. 69–75, 2011, doi: 10.1016/j.nuclphysa.2011.06.032.

[16] S. N. Gupta, S. F. Radford, and W. W. Repko, "Semirelativistic Potential

Model for Heavy Quarkonia," *Phys.\ Rev.\ D*, vol. 34, pp. 201–206, 1986, doi: 10.1103/PhysRevD.34.201.

[17] J. N. Pandya and P. C. Vinodkumar, "Masses of S and P wave mesons and pseudoscalar decay constants using a confinement scheme," *Pramana - J. Phys.*, vol. 57, no. 4, pp. 821–827, 2001, doi: 10.1007/s12043-001-0031-y.

[18] J. N. Pandya, N. R. Soni, N. Devlani, and A. K. Rai, "Decay rates and electromagnetic transitions of heavy quarkonia," *Chin.\ Phys.\ C*, vol. 39, no. 12, p. 123101, 2015, doi: 10.1088/1674-1137/39/12/123101.

[19] A. K. Rai, J. N. Pandya, and P. C. Vinodkumar, "Decay rates of quarkonia with NRQCD formalism using spectroscopic parameters of potential models," *Eur. Phys. J. A*, vol. 38, no. 1, pp. 77–84, 2008, doi: 10.1140/epja/i2008-10639-9.

[20] A. K. Rai, R. H. Parmar, and P. C. Vinodkumar, "Masses and decay constants of heavy-light flavor mesons in a variational scheme," *J.\ Phys.\ G*, vol. 28, pp. 2275–2282, 2002, doi: 10.1088/0954-3899/28/8/313.

[21] A. K. Rai, B. Patel, and P. C. Vinodkumar, "Properties of $Q\bar{Q}$ mesons in nonrelativistic QCD formalism," *Phys. Rev. C - Nucl. Phys.*, vol. 78, no. 5, p. 055202, Nov. 2008, doi: 10.1103/PhysRevC.78.055202.

[22] B. Patel and P. C. Vinodkumar, "Properties of QQ(Q ∈ b, c) mesons in Coulomb plus power potential (CPPv)," *J. Phys. G Nucl. Part. Phys.*, vol. 36, no. 3, 2009, doi: 10.1088/0954-3899/36/3/035003.

[23] V. Sauli, "Bethe-Salpeter study of radially excited vector quarkonia," *Phys. Rev. D - Part. Fields, Gravit. Cosmol.*, vol. 86, no. 9, p. 096004, Nov. 2012, doi: 10.1103/PhysRevD.86.096004.

[24] C. S. Fischer, S. Kubrak, and R. Williams, "Spectra of heavy mesons in the Bethe-Salpeter approach," *Eur. Phys. J. A*, vol. 51, no. 1, pp. 1–9, 2015, doi: 10.1140/epja/i2015-15010-7.

[25] S. Leitão, A. Stadler, M. T. Peña, and E. P. Biernat, "Linear confinement in momentum space: Singularity-free bound-state equations," *Phys. Rev. D - Part. Fields, Gravit. Cosmol.*, vol. 90, no. 9, p. 096003, Nov. 2014, doi: 10.1103/PhysRevD.90.096003.

[26] C. Berkdemir, A. Berkdemir, and R. Sever, "Polynomial solutions of the Schrödinger equation for the generalized Woods-Saxon potential," *Phys. Rev. C - Nucl. Phys.*, vol. 72, no. 2, p. 027001, Aug. 2005, doi: 10.1103/PhysRevC.72.027001.

[27] B. J. Falaye, K. J. Oyewumi, and M. Abbas, "Exact solution of Schrödinger equation with q-deformed quantum potentials using Nikiforov - Uvarov method," *Chinese Phys. B*, vol. 22, no. 11, 2013, doi: 10.1088/1674-1056/22/11/110301.

[28] B. I. Ita and A. I. Ikeuba, "Solutions of the Dirac Equation with Gravitational plus Exponential Potential," *Appl. Math.*, vol. 04, no. 10, pp. 1–6, 2013, doi:

10.4236/am.2013.410a3001.

[29]  F. Yaşuk, C. Berkdemir, and A. Berkdemir, "Exact solutions of the Schrödinger equation with non-central potential by the Nikiforov-Uvarov method," *J. Phys. A. Math. Gen.*, vol. 38, no. 29, pp. 6579–6586, 2005, doi: 10.1088/0305-4470/38/29/012.

[30]  H. Karayer, D. Demirhan, and F. Büyükkılıç, "Conformable Fractional Nikiforov - Uvarov Method," *Commun. Theor. Phys.*, vol. 66, no. 1, pp. 12–18, 2016, doi: 10.1088/0253-6102/66/1/012.

[31]  A. F. Al-Jamel and H. Widyan, "Heavy Quarkonium Mass Spectra in A Coulomb Field Plus Quadratic Potential Using Nikiforov-Uvarov Method," *Appl. Phys. Res.*, vol. 4, no. 3, pp. 94–99, 2012, doi: 10.5539/apr.v4n3p94.

[32]  S. M. Ikhdair, " Approximate l -States of the Manning-Rosen Potential by Using Nikiforov-Uvarov Method ," *ISRN Math. Phys.*, vol. 2012, pp. 1–20, 2012, doi: 10.5402/2012/201525.

[33]  B. I. Ita, C. O. Ehi-Eromosele, A. Edobor-Osoh, A. I. Ikeuba, and T. A. Anake, "Solutions of the Schrödinger equation with inversely quadratic effective plus Mie-type potential using Nikiforov-Uvarov method," in *AIP Conference Proceedings*, 2014, vol. 1629, no. November, pp. 235–238, doi: 10.1063/1.4902278.

[34]  B. Ita, P. Tchoua, E. Siryabe, and G. E. Ntamack, "Solutions of the Klein-Gordon Equation with the Hulthen Potential Using the Frobenius Method," vol. 4, no. 5, pp. 173–177, 2014, doi: 10.5923/j.ijtmp.20140405.02.

[35]  W. Kwong and J. L. Rosner, "D-wave quarkonium levels of the family," *Phys. Rev. D*, vol. 38, no. 1, pp. 279–297, 1988, doi: 10.1103/PhysRevD.38.279.

[36]  M. B. Voloshin, "Charmonium," *Progress in Particle and Nuclear Physics*, vol. 61, no. 2. pp. 455–511, 2008, doi: 10.1016/j.ppnp.2008.02.001.

[37]  W. J. Deng, H. Liu, L. C. Gui, and X. H. Zhong, "Charmonium spectrum and electromagnetic transitions with higher multipole contributions," *Phys. Rev. D*, vol. 95, no. 3, pp. 1–20, 2017, doi: 10.1103/PhysRevD.95.034026.

[38]  P. P. D'Souza, A. P. Monteiro, and K. B. V. Kumar, "Properties of Low-Lying Charmonium States in a Phenomenological Approach," pp. 1–10, 2017.

[39]  I. Haysak, Y. Fekete, V. Morokhovych, S. Chalupka, and M. Salak, "Spin effects in two quark system and mixed states," *Czechoslov. J. Phys.*, vol. 55, no. 5, pp. 541–554, 2005, doi: 10.1007/s10582-005-0059-1.

[40]  T. Barnes, S. Godfrey, and E. S. Swanson, "Higher charmonia," *Phys. Rev. D - Part. Fields, Gravit. Cosmol.*, vol. 72, no. 5, pp. 1–28, 2005, doi: 10.1103/PhysRevD.72.054026.

[41]  O. Lakhina and E. S. Swanson, "Dynamic properties of charmonium," *Phys. Rev. D - Part. Fields, Gravit. Cosmol.*, vol. 74, no. 1, pp. 1–18, 2006, doi: 10.1103/PhysRevD.74.014012.

[42] L. Motyka and K. Zalewski, "Mass spectra and leptonic decay widths of heavy quarkonia," *Eur. Phys. J. C*, vol. 4, no. 1, pp. 107–114, 1998, doi: 10.1007/s100529800743.

[43] E. Eichten and F. Feinberg, "Spin-dependent forces in quantum chromodynamics," *Phys. Rev. D*, vol. 23, no. 11, pp. 2724–2744, 1981, doi: 10.1103/PhysRevD.23.2724.

[44] C. Patrignani *et al.*, "Review of particle physics," *Chinese Physics C*, vol. 40, no. 10. 2016, doi: 10.1088/1674-1137/40/10/100001.

[45] H. Mansour and A. Gamal, "Bound State of Heavy Quarks Using a General Polynomial Potential," *Adv. High Energy Phys.*, vol. 2018, no. 2, 2018, doi: 10.1155/2018/7269657.

[46] D. Ebert, R. N. Faustov, and V. O. Galkin, "Spectroscopy and Regge trajectories of heavy quarkonia and Bc mesons," *Eur. Phys. J. C*, vol. 71, no. 12, pp. 1–13, 2011, doi: 10.1140/epjc/s10052-011-1825-9.

[47] N. R. Soni, B. R. Joshi, R. P. Shah, H. R. Chauhan, and J. N. Pandya, "$Q\bar{Q}$ ($Q \in \{b, c\}$) spectroscopy using the Cornell potential," *Eur. Phys. J. C*, vol. 78, no. 7, 2018, doi: 10.1140/epjc/s10052-018-6068-6.

[48] B. Q. Li and K. T. Chao, "Higher charmonia and X, Y, Z states with screened potential," *Phys. Rev. D - Part. Fields, Gravit. Cosmol.*, vol. 79, no. 9, 2009, doi: 10.1103/PhysRevD.79.094004.

[49] M. Shah, A. Parmar, and P. C. Vinodkumar, "Leptonic and digamma decay properties of S-wave quarkonia states," *Phys. Rev. D - Part. Fields, Gravit. Cosmol.*, vol. 86, no. 3, pp. 4–7, 2012, doi: 10.1103/PhysRevD.86.034015.

[50] O. Lakhina and E. S. Swanson, "Dynamic properties of charmonium," *Phys. Rev. D - Part. Fields, Gravit. Cosmol.*, vol. 74, no. 1, 2006, doi: 10.1103/PhysRevD.74.014012.

[51] S. Patel, P. C. Vinodkumar, and S. Bhatnagar, "Decay rates of charmonia within a quark-antiquark confining potential," *Chinese Phys. C*, vol. 40, no. 5, 2016, doi: 10.1088/1674-1137/40/5/053102.

[52] S. Godfrey and K. Moats, "Bottomonium mesons and strategies for their observation," *Phys. Rev. D*, vol. 92, no. 5, p. 054034, Sep. 2015, doi: 10.1103/PhysRevD.92.054034.

[53] B. Q. Li and K. T. Chao, "Bottomonium spectrum with screened potential," *Commun. Theor. Phys.*, vol. 52, no. 4, pp. 653–661, 2009, doi: 10.1088/0253-6102/52/4/20.

[54] J. Segovia, P. G. Ortega, D. R. Entem, and F. Fernández, "Bottomonium spectrum revisited," *Phys. Rev. D*, vol. 93, no. 7, 2016, doi: 10.1103/PhysRevD.93.074027.

[55] N. Devlani, V. Kher, and A. K. Rai, "Masses and electromagnetic transitions of

the Bc mesons," *Eur. Phys. J. A*, vol. 50, no. 10, pp. 627–631, 2014, doi: 10.1140/epja/i2014-14154-2.

[56] S. Godfrey, "Spectroscopy of B c mesons in the relativized quark model," *Phys. Rev. D*, vol. 70, no. 5, p. 054017, Sep. 2004, doi: 10.1103/PhysRevD.70.054017.

[57] A. P. Monteiro, M. Bhat, and K. B. V. Kumar, "Cb̄ Spectrum and Decay Properties With Coupled Channel Effects," *Phys. Rev. D*, vol. 95, no. 5, pp. 1–12, 2017, doi: 10.1103/PhysRevD.95.054016.

# Appendix

This is a flow chart about how one can calculate the theoretical data. It consists of three stages to do that:

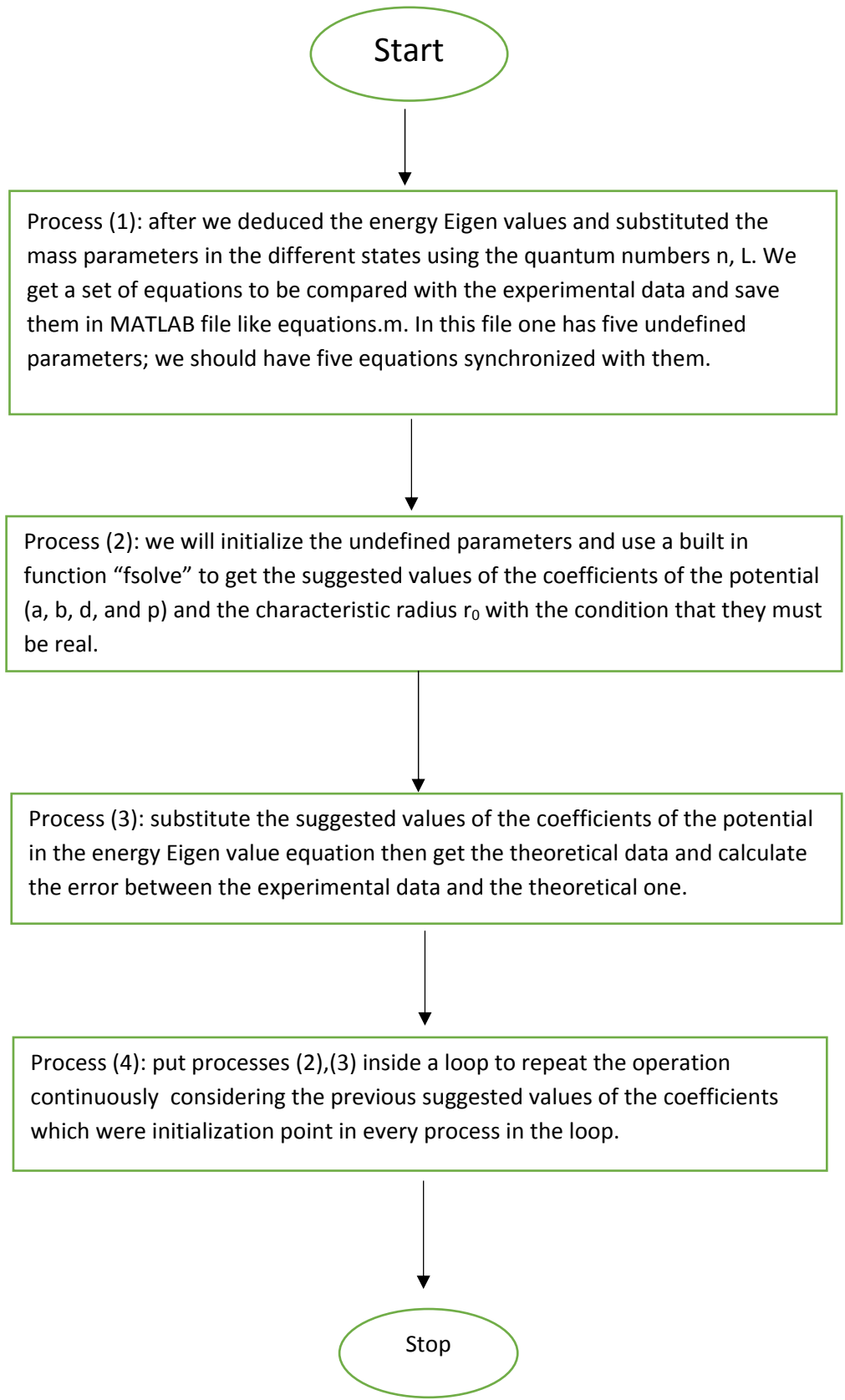

```
                    ┌─────────┐
                    │  Start  │
                    └─────────┘
                         │
                         ▼
```

Process (1): after we deduced the energy Eigen values and substituted the mass parameters in the different states using the quantum numbers n, L. We get a set of equations to be compared with the experimental data and save them in MATLAB file like equations.m. In this file one has five undefined parameters; we should have five equations synchronized with them.

Process (2): we will initialize the undefined parameters and use a built in function "fsolve" to get the suggested values of the coefficients of the potential (a, b, d, and p) and the characteristic radius $r_0$ with the condition that they must be real.

Process (3): substitute the suggested values of the coefficients of the potential in the energy Eigen value equation then get the theoretical data and calculate the error between the experimental data and the theoretical one.

Process (4): put processes (2),(3) inside a loop to repeat the operation continuously considering the previous suggested values of the coefficients which were initialization point in every process in the loop.

```
                    ┌─────────┐
                    │  Stop   │
                    └─────────┘
```

We found that after some iteration processes in the loops that, the previous suggested values of the coefficients are the same in all last processes, as there is no change in the data. So, we used another program which gives a small change in the parameters values which was Origin lab program.

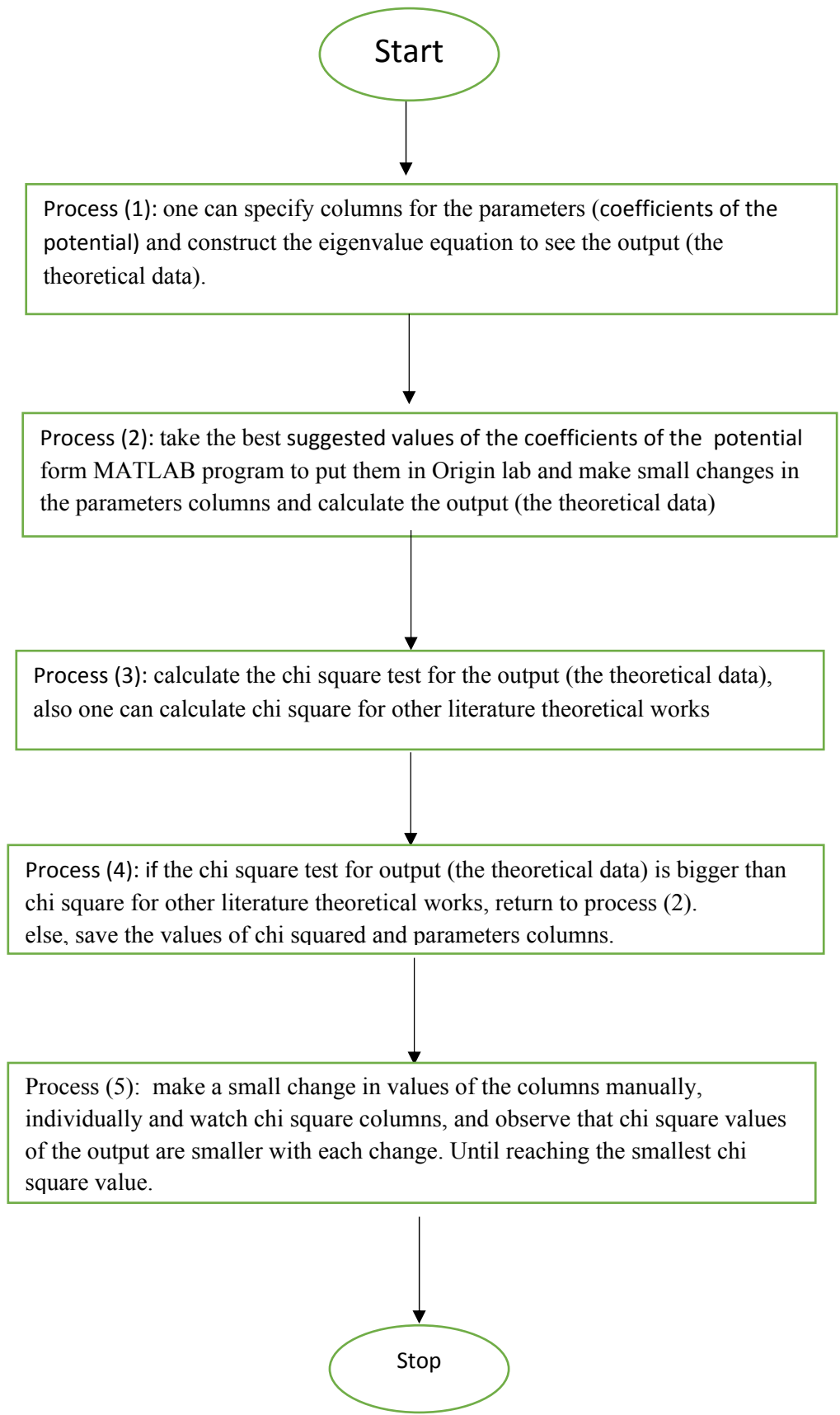

Third stage, after we got the values of the coefficients of the potential with the smallest chi square values from the origin program, we will substitute them in the energy eigenvalue equation and calculate the theoretical data (present work) for the different n, L states by using "fsolve" built in function in MATLAB like first flow chart.